\documentclass[12pt,preprint]{aastex}

\slugcomment{Accepted by the Astrophysical Journal}
\shorttitle{Turbulent Mixing in the Outer Solar Nebula}
\shortauthors{Turner et al.}

\begin{document}
\title{Turbulent Mixing in the Outer Solar Nebula}

\author{N. J. Turner, K. Willacy, G. Bryden and H. W. Yorke}

\affil{Jet Propulsion Laboratory MS 169-506, California Institute of
Technology, Pasadena CA 91109; neal.turner@jpl.nasa.gov}

\begin{abstract}
The effects of turbulence on the mixing of gases and dust in the outer
Solar nebula are examined using 3-D MHD calculations in the
shearing-box approximation with vertical stratification.  The
turbulence is driven by the magneto-rotational instability.  The
magnetic and hydrodynamic stresses in the turbulence correspond to an
accretion time at the midplane about equal to the lifetimes of T~Tauri
disks, while accretion in the surface layers is thirty times faster.
The mixing resulting from the turbulence is also fastest in the
surface layers.  The mixing rate is similar to the rate of radial
exchange of orbital angular momentum, so that the Schmidt number is
near unity.  The vertical spreading of a trace species is well-matched
by solutions of a damped wave equation when the flow is
horizontally-averaged.  The damped wave description can be used to
inexpensively treat mixing in 1-D chemical models.  However, even in
calculations reaching a statistical steady state, the concentration at
any given time varies substantially over horizontal planes, due to
fluctuations in the rate and direction of the transport.  In addition
to mixing species that are formed under widely varying conditions, the
turbulence intermittently forces the nebula away from local chemical
equilibrium.  The different transport rates in the surface layers and
interior may affect estimates of the grain evolution and molecular
abundances during the formation of the Solar system.
\end{abstract}


\section{INTRODUCTION\label{sec:intro}}

The compositions of the gas giant planets and the comets reflect the
conditions that were found in the outer Solar nebula.  Jupiter,
Saturn, Uranus and Neptune are enriched in volatiles relative to the
Sun \citep{lo00,yo03}.  The volatiles may have been accreted on the
planets while trapped in ices \citep{ls85,bk88,hg04}, but the
efficiency of this process depends on the poorly-known molecular
compositions of the protoplanetary feeding zones, which depend in turn
on the chemical reactions at work in different parts of the nebula and
the effectiveness of transport by accretion and mixing.  Some comets
contain crystalline silicate grains \citep{wh99} and some have HCN
molecules with deuterium to hydrogen ratios intermediate between the
Sun and molecular clouds \citep{lo00}.  These results are puzzling
because the midplane temperatures where the comets formed, near 30~AU,
were too low for either the annealing of silicates or the destruction
of primordial ices.  The anomalous compositions of the comets could be
due to gas and dust transported from the surface layers of the nebula
or from nearer the Sun, where the temperatures were higher.

The molecular makeup of the Solar nebula was profoundly affected by
chemical reactions in the gas phase and on grain surfaces, by ionizing
radiation from the young Sun \citep{wl00} and by the mixing of
materials formed at hot and cold locations \citep{mv84,ga01}.
Measurements of the millimeter molecular line fluxes in present-day
protostellar disks are translated into abundances using detailed
models in which the rates of transfer of material between different
regions are a major uncertainty.  Calculations without turbulent
mixing indicate a layered structure in the outer nebula due to
ultraviolet surface irradiation and interior freeze-out \citep{av02}.
Molecules such as CO and N$_2$ form from photodissociation products in
warm surface layers, and freeze on grain surfaces near the midplane
\citep{cd05}.  Including vertical turbulent mixing leads to
significant changes in the chemical composition in models spanning
radial distances 1--10~AU \citep{ih04} as well as outside 100~AU
\citep{wl06}.  The turbulence in these chemical models is assumed to
act like a diffusion process.  The importance of the mixing relative
to the overall accretion flow toward the young star depends on the
ratio between the material diffusion and angular momentum transfer
coefficients \citep{st90}, typically assumed to be unity.

The growth of solid bodies in protostellar disks depends fundamentally
on the motions of the gas.  Interstellar grains can grow larger than
30~$\mu$m in $10^4$~yr through collisions in turbulence of moderate
strength \citep{sy01}.  Small grains are removed so efficiently by
coagulation that the overall infrared excess should fall below
observed values long before the disk reaches an age of $10^6$~yr
typical of T~Tauri stars \citep{dd05}.  This problem may be resolved
if the small particles are replenished by the break-up of aggregate
grains in high-speed collisions.  Many models of the particle size
evolution include collisions resulting from a parametrized turbulence.
The velocity fluctuations are often chosen to be an arbitrary fixed
fraction of the sound speed and independent of height at each distance
from the star.  Better characterization of the turbulence is needed to
further understand how the solid bodies are assembled.

A possible driver of turbulence in the Solar nebula is the
magneto-rotational instability or MRI \citep{bh91}.  The MRI occurs
where weak magnetic fields are present and sufficiently coupled to the
orbiting gas.  Magnetic fields are detected in the molecular cloud
cores that collapse to form protostars \citep{tc96} and in the disks
around T~Tauri stars \citep{th99}.  Remanent magnetization is found in
meteoroids that solidified early in the history of the Solar system
\citep{ch91}.  The coupling between the disk gas and the fields
depends on the conductivity.  The outer Solar nebula was too cool for
thermal ionization, and the conductivity is determined by balancing
recombination on grain surfaces with the ionization due to cosmic rays
\citep{wn99} or X-rays \citep{in05}.  We are interested here in the
region near the present orbit of Neptune, where the conductivity
models indicate the MRI was active for reasonable grain parameters
\citep{sm00}.  The instability leads to turbulence in which accretion
is driven by the transfer of orbital angular momentum outward along
magnetic field lines \citep{hg95}.  The mean accretion stress due to
the magnetic forces is several times greater than that due to gas
pressure gradients.  The magnetic fields are anisotropic, with the
azimuthal component strongest because of the differential rotation,
the radial component next due to the radial stretching of field lines
during the angular momentum transfer, and the vertical component
weakest.  The fields are buoyant, leading to the formation of a
magnetized corona above and below the disk in calculations including
the vertical stratification \citep{ms00}.  The velocity dispersion in
the corona is greater than in the disk interior.

Accretion and mixing act together to redistribute material through
protostellar disks.  Existing results on the relative importance of
these two processes in MRI turbulence are as follows.  The angular
momentum transfer coefficient is an order of magnitude greater than
the turbulent diffusion coefficient of a passive tracer species, in
calculations with a net vertical magnetic field by \cite{cs05} using
the {\sc Zeus} MHD code.  On the other hand, the mixing of a dust
fluid coupled to the gas by friction forces occurs at rates similar to
the accretion in calculations with zero net magnetic flux by
\cite{jk05} using the {\sc Pencil} code.  Both studies involve
unstratified shearing-box calculations.  Our work resembles
\cite{cs05} in following a passive tracer with the {\sc Zeus} code.
We consider both zero and non-zero magnetic fluxes and include the
effects of vertical stratification.  The equations and method of
solution are described in section~\ref{sec:eqns}, the domain and
initial conditions in section~\ref{sec:ic}.  The results are divided
into two sections.  The spreading of an initial concentration
enhancement in the turbulence is compared with solutions of a damped
wave equation in section~\ref{sec:mixing}.  The approximate steady
state resulting from transport of material away from a fixed source
and the chemical histories of individual Lagrangian fluid elements are
discussed in section~\ref{sec:steady}.  A summary and conclusions are
in section~\ref{sec:conc}.

\section{EQUATIONS AND METHOD OF SOLUTION\label{sec:eqns}}

The domain is a small patch of the disk, centered at the midplane a
distance $R$ from the axis of rotation.  The effects of the radial
gradient in orbital angular speed are treated using the local shearing
box approximation \citep{hg95} with vertical stratification included.
Curvature along the direction of orbital motion is neglected, and the
resulting Cartesian coordinate system co-rotates at the Keplerian
orbital frequency $\Omega = \left(GM/R^3\right)^{1/2}$ for domain
center with a stellar mass $M$.  Coriolis and tidal forces in the
rotating frame and the component of gravity perpendicular to the
midplane are all included.  The local coordinates $(x, y, z)$
correspond to distance from the domain center along the radial,
orbital, and vertical directions, respectively.  The azimuthal
boundaries are periodic and the radial boundaries are
shearing-periodic.  Fluid passing through one radial boundary
reappears on the other at an azimuth which varies in time according to
the difference in orbital speed across the box.  The difference is
computed using a Keplerian profile linearized about domain center.
The vertical boundaries are open and allow gas to flow out but not in.

The equations of isothermal ideal magnetohydrodynamics (MHD) are
solved in the co-rotating frame.  Conservation of mass and momentum
and the evolution of magnetic fields are described by
\begin{equation}\label{eqn:cty}
{{\partial\rho}\over{\partial t}}+{\bf\nabla\cdot}(\rho{\bf v})=0,
\end{equation}
\begin{equation}\label{eqn:eomg}
{{\partial{\bf v}}\over{\partial t}} + {\bf v \cdot \nabla v} =
	- {{\bf\nabla}p\over\rho}
	+ {1\over 4\pi\rho}({\bf\nabla\times B}){\bf\times B}
	- 2{\bf\Omega\times v} + 3\Omega^2 {\bf\hat x}
	- \Omega^2 {\bf\hat z},
\end{equation}
and
\begin{equation}\label{eqn:dbdt}
{\partial{\bf B}\over\partial t} = {\bf\nabla\times}({\bf v\times B}),
\end{equation}
together with the isothermal equation of state $p=c_s^2\rho$, which
depends on the isothermal sound speed, $c_s$.  The density, velocity,
gas pressure and magnetic field are $\rho$, ${\bf v}$, $p$ and ${\bf
B}$, respectively.  The equations are integrated using the standard
{\sc Zeus} code \citep{sn92a,sn92b,hs95}.  The flow of a trace
contaminant is followed by solving an additional continuity equation
\begin{equation}\label{eqn:color}
{{\partial C}\over{\partial t}}+{\bf\nabla\cdot}(C{\bf v})=0
\end{equation}
for the volume mass density, $C$.

We also wish to track the paths of individual fluid elements through
the flow.  Advanced trajectory integration methods offer few benefits
here because MRI-driven turbulence is chaotic and the streamlines
diverge exponentially with an $e$-folding time similar to the orbital
period \citep{wb03}.  An appropriate choice is an integration method
that is slightly more accurate than the hydrodynamic part of the
calculations.  We use a classical fourth-order Runge-Kutta particle
orbit integrator.  Gas quantities are interpolated to the particle
positions using the same second-order accuracy as in the
hydrodynamics.  Tests show the method has the expected order of
convergence.  In one example, a washing-machine problem, a circulating
velocity field oscillates in time.  The error in the particle position
has the $(\Delta t)^4$ dependence of the fourth-order method and is
dominated by roundoff error when the number of timesteps per
oscillation is large.

\section{DOMAIN AND INITIAL CONDITIONS\label{sec:ic}}

The surface density and temperature for the calculations are chosen
based on observations of disks around T~Tauri stars.  The domain is
centered 30~AU from an object of one Solar mass so that the orbital
period $2\pi/\Omega = 164$~years.  The surface density $\Sigma=25$
g~cm$^{-2}$ is consistent with fits to the spectral energy
distributions of young stars \citep{dc01}.  The equilibrium
temperature of a blackbody 30~AU from the Sun is 50~K, but detailed
radiative transfer calculations show the midplane of the Solar nebula
was colder than a blackbody in direct sunlight, due to the obscuration
by the inner disk.  For a T~Tauri star of 0.5~Solar mass and 0.9~Solar
luminosity, the midplane temperature at 30~AU is estimated to be about
25~K \citep{dc01}.  Temperatures above and below the midplane are
greater because the material is externally-heated by thermal emission
from grains in surface layers exposed to the stellar irradiation
\citep{cg97}.  For our calculations spanning the midplane we use an
isothermal equation of state with temperature $T=25$~K throughout.
The optical depth of the disk to its own thermal radiation is
about~0.05, so that internal temperature fluctuations are quickly
erased by radiation losses.

The initial state is in radial and vertical hydrostatic equilibrium.
The density varies with height as $\rho=\rho_0 \exp({-z^2/2H^2})$
having scale height $H=c_s/\Omega=1.64$~AU.  The uniform sound speed
$c_s=0.298$ km s$^{-1}$ is appropriate for a mean molecular weight
$\mu=2.34$, while the midplane density $\rho_0=4.05\times 10^{-13}$
g~cm$^{-3}$ is chosen to give the desired surface mass density.  The
resulting structure is locally gravitationally stable, as the Toomre
parameter $Q=\Omega c_s/(\pi G \Sigma)=7$.  Small dust particles are
well-coupled to the gas throughout the domain by the Epstein drag
force.  Compact spherical ice grains less than 100~$\mu$m in radius
have any motions with respect to the gas damped within 0.1~orbits even
at the lowest gas densities.

The coupling between gas and magnetic fields in the outer Solar nebula
depends on the balance between recombination on grain surfaces and
ionization by cosmic rays and X-rays.  The conductivity varies with
the total grain surface area per unit gas mass \citep{wn99}.  For an
interstellar particle size distribution, the surface area lies mostly
on the smallest grains of about 0.005~$\mu$m \citep{dl84}.  Under the
conditions at the midplane in our calculations, the free charge
resides largely on the grains, the ionization fraction is very low,
and the magneto-rotational instability is absent \citep{sm00}.
However if the smallest grains are agglomerated into larger bodies and
the remaining surface area is dominated by 0.1-$\mu$m grains, the
largest contributions to the conductivity are from electrons and ions
\citep{wn99}.  The conductivity is ample for MRI \citep{sm00}, but
ambipolar diffusion may be present.  Ambipolar diffusion can lead to
weaker turbulence \citep{hs98}.  If the grain surface area is reduced
to negligible levels and the magnetic pressure is much less than the
gas pressure, the Hall term is the largest non-ideal effect
\citep{ss02a,sw05}.  The magnetic field is tied to the gas and the
instability leads to turbulence with properties similar to those in
ideal MHD \citep{ss02b}.  Given the rapid removal of small grains by
coagulation in the existing models \citep{dd05} and the observational
evidence for grains in disks larger than in the interstellar medium
\citep{cw01,dc01,md03,sb03}, we consider the case where gas and fields
are well-coupled.

The initial $y$-velocities in our calculations are chosen for radial
force balance between gravity and rotation.  The $x$- and
$z$-velocities in each grid zone are initially set to random values
with maximum magnitudes 1\% of the sound speed.  The domain size in
the radial, azimuthal and vertical directions is $L_x\times L_y\times
L_z = 3\times 12\times 12$~AU and the grid extends $3.7H$ either side
of the midplane.  The standard resolution is $32\times 64\times 128$
grid zones, corresponding to a zone aspect ratio $1:2:1$.  An
additional calculation is made with the spatial resolution doubled
giving twice as many zones along each direction.

The magnetic field has zero net flux.  The vertical component is
initially proportional to $\sin\left(2\pi x/L_x\right)$ and the
azimuthal component to $\cos\left(2\pi x/L_x\right)$, so that the
field lines are straight, the magnetic pressure is uniform and there
are no magnetic forces.  The initial field strength $4.9\times
10^{-3}$~Gauss corresponds to a magnetic pressure 379~times smaller
than the midplane gas pressure $\rho_0 c_s^2$.  The wavelength $2\pi
v_A/\Omega$ of the magneto-rotational instability at the midplane is
eight times the vertical grid spacing or $0.45 H$.  During the
calculations, large Alfv\'{e}n speeds $v_A$ and small timesteps are
prevented by imposing a density floor $10^{-18}$ g cm$^{-3}$ that lies
in the range expected for material falling in from a surrounding
protostellar envelope.  A check on the effects of the density floor is
described in section~\ref{sec:boundaries}.  The field on the vertical
boundaries is evolved by applying the condition $\partial{\bf\cal
E}/\partial z = 0$ to the electromotive force ${\bf\cal E} \equiv {\bf
v\times B}$.

At the start of each calculation, the magneto-rotational instability
grows to non-linear amplitudes over about two orbits.  The flow then
becomes turbulent and after four orbits the total kinetic energy
varies little with time.  At ten orbits, the magnetic field is
thoroughly tangled and the flow has reached a time-averaged steady
state.  A contaminant is then added and its motions tracked.

\section{SPREADING OF AN INITIAL CONCENTRATION ENHANCEMENT\label{sec:mixing}}

\begin{figure}[tb]
\epsscale{0.4}
\plotone{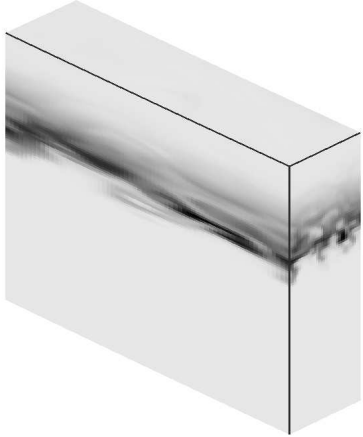}
\caption{\small Spreading of a layer of contaminant in a 3-D MHD
calculation of the turbulence driven by MRI.  The layer is added at
ten orbits, when the turbulence is well-developed, and results are
shown five orbits later.  The gray levels indicate the contaminant
density on the domain walls.  The darkest shades correspond to 37\% of
the initial value.  The radial coordinate increases toward top right,
the azimuthal coordinate toward top left.  An {\sc MPEG} animation in
the online edition shows the evolution from 10 to 20~orbits.  The
contaminant density color scale in the animation is logarithmic
between $\leq$1\% (black) and $\geq$100\% of the initial value
(white).
\label{fig:csnapshot}}
\end{figure}

In the fiducial calculation, the contaminant is initially placed in a
thin horizontal layer, centered 3~AU above the midplane and 0.375~AU
thick.  The density of the contaminant is set to $C_0$ in the layer
and zero elsewhere.  The contaminant spreads irregularly due to the
turbulent motions, as shown in figure~\ref{fig:csnapshot} by a
snapshot at 15~orbits.

We wish to describe the turbulent mixing using a simple
one-dimensional equation.  The result is most easily written in terms
of the concentration, or contaminant per unit mass, $Q=C/\rho$.  The
concentration is split into mean and fluctuating parts,
\begin{equation}
Q = {\bar Q} + q,
\end{equation}
where ${\bar Q}$ is averaged over the horizontal extent of the domain,
\begin{equation}
{\bar Q} = {1\over L_x L_y}\int\int Q\,dx dy.
\end{equation}
A similar horizontal average of the velocity is zero since there is no
overall expansion or contraction of the disk.  Writing the flux of the
contaminant ${\bf F}=Q{\bf v}$ and its horizontal average ${\bf\bar
F}=\overline{q{\bf v}}$, the contaminant continuity equation is
\begin{equation}\label{eqn:qcty}
{\partial Q\over\partial t} = -{\bf\nabla\cdot F} + Q{\bf\nabla\cdot v}
\end{equation}
and its mean
\begin{equation}\label{eqn:qbarcty}
{\partial{\bar Q}\over\partial t} =
-{\bf\nabla\cdot{\bar F}}.
\end{equation}
Extra terms in equation~\ref{eqn:qbarcty} are zero because the
time-steady mean density profile implies ${\overline{\bf\nabla\cdot
v}}=0$, while ${\overline{q{\bf\nabla\cdot v}}}=0$ since $q$ and
${\bf\nabla\cdot v}$ are uncorrelated and both have zero mean.  We now
construct a damped wave equation \citep{bf03,bk04} to avoid the
infinite signal speeds associated with a diffusion description.  The
mean continuity equation is divided by a characteristic time $\tau$
and added to its time derivative, yielding
\begin{equation}\label{eqn:tauddt}
{\partial^2 {\bar Q} \over \partial t^2}
 + {1\over\tau} {\partial {\bar Q} \over \partial t}
 = -{\bf\nabla\cdot} \left({\partial{\bf\bar F}\over\partial t}
     + {{\bf\bar F}\over\tau}\right).
\end{equation}
To obtain an equation in ${\bar Q}$, we eliminate fluctuating
quantities.  The time-derivative of the flux has a term
$\overline{q{\dot{\bf v}}}$ that vanishes since the acceleration and
the density are uncorrelated and each averages to zero.  The remaining
term $\overline{{\dot q}{\bf v}}$ is evaluated using ${\dot q} = {\dot
Q}-{\dot{\bar Q}} = -{\bf\nabla\cdot F} + Q{\bf\nabla\cdot v} +
{\bf\nabla\cdot{\bar F}}$, leading to
\begin{equation}
{\partial{\bf\bar F}\over\partial t} =
 - \overline{{\bf v v\cdot\nabla}q}
 - \overline{{\bf v v}}{\bf\cdot}{\bf\nabla}{\bar Q}.
\end{equation}
The first term on the right is a triple-correlation of fluctuating
quantities and can be approximated by the ratio of the flux to the
characteristic time, $-{\bf\bar F}/\tau$ \citep{bf03}.  The evolution
equation~\ref{eqn:tauddt} for the horizontally-averaged contaminant
density then takes the form of a damped wave equation,
\begin{equation}\label{eqn:dampedwave}
{\partial^2 {\bar Q} \over \partial t^2}
 + {1\over\tau} {\partial {\bar Q} \over \partial t}
 = {\partial \over \partial z} \left({\overline{v_z^2}}
   {\partial {\bar Q} \over \partial z}\right).
\end{equation}
Equation~\ref{eqn:dampedwave} is a diffusion equation with a second
time derivative term added on the left to make a damped wave equation.
The diffusion coefficient is the product of the characteristic time
and the squared velocity dispersion, so that we can identify the
characteristic time as the correlation time of the turbulence.
Including the wave term gives the desirable property that the
transport is no faster than the RMS turbulent speed.

\begin{figure}[tb]
\epsscale{0.4}
\plotone{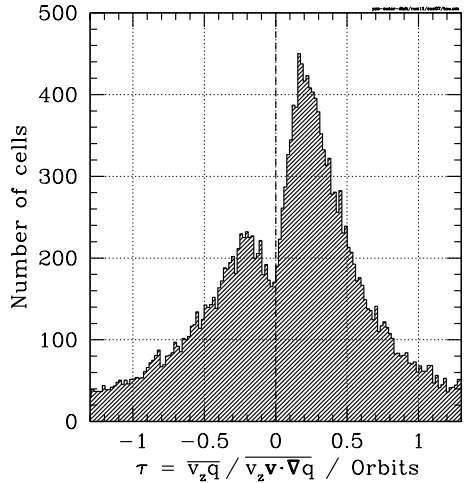}
\caption{\small Histogram of turbulent correlation times $\tau$
measured in the MHD calculation between 10 and 30~orbits.  The
measurements are made every 0.1~orbits at each height on the
computational grid.  Zero correlation time, indicated by a vertical
dashed line, occurs less often than small positive and negative
values.  The median correlation time is 0.163~orbits.
\label{fig:tau}}
\end{figure}

\begin{figure}[tb]
\epsscale{0.4}
\plotone{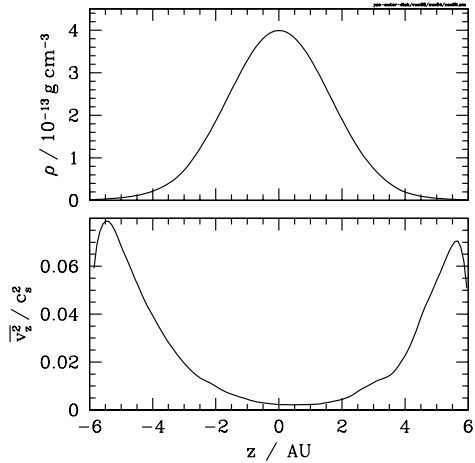}
\caption{\small Density profile (top) and vertical velocity dispersion
(bottom) in the MHD calculation.  Both are averaged horizontally and
over time from 10 to 30~orbits, and are plotted against height.  The
velocity dispersion is scaled using the isothermal sound speed $c_s$.
\label{fig:coefft}}
\end{figure}

\begin{figure}[tb]
\epsscale{0.8}
\plotone{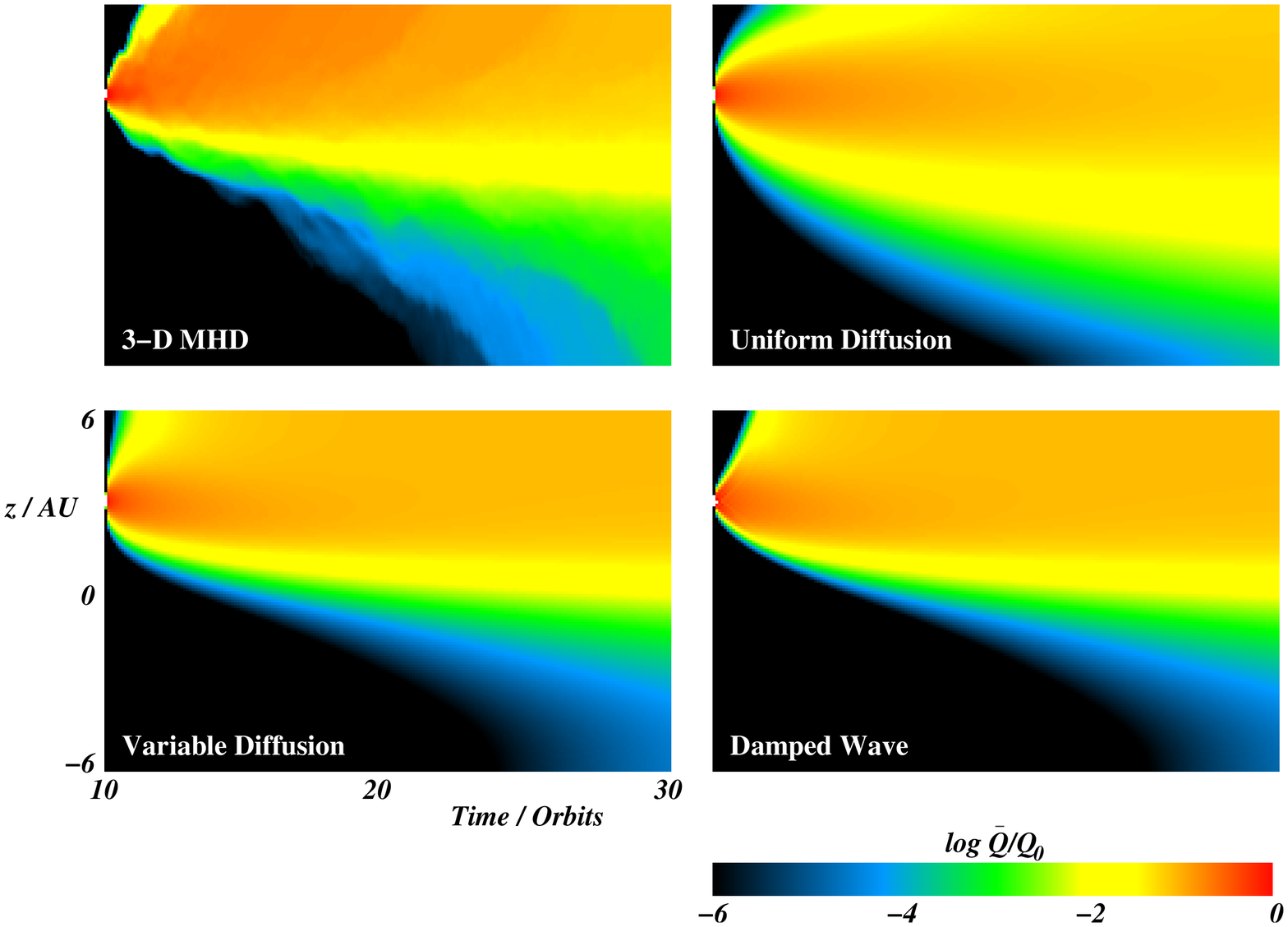}
\caption{\small Horizontally-averaged contaminant concentration $\bar
Q$ (colors) versus height and time in the MHD calculation (top left)
and three models.  The solution of the diffusion equation with uniform
coefficient is shown at top right, the diffusion equation with
height-dependent coefficient at bottom left, and the damped wave
equation~\ref{eqn:dampedwave} at bottom right.  In the uniform
diffusion case the contaminant spreads too quickly in the interior and
too slowly in the surface layers.  In the variable diffusion case the
contaminant spreads too quickly in the first few orbits.  The best
match is with the full damped wave solution.
\label{fig:allterms}}
\end{figure}

The two parameters in equation~\ref{eqn:dampedwave} are readily
measured in the MHD calculation.  The correlation time $\tau={\bar
F_z}/(\overline{v_z {\bf v\cdot\nabla}q})$ between 10 and 30~orbits
has a median value of 0.163~orbits, comparable to the shear time
$(\frac{3}{2}\Omega)^{-1} = 0.106$~orbits.  The range of values is
rather wide and is shown in figure~\ref{fig:tau}.  The RMS vertical
speed $(\overline{v_z^2})^{1/2}$ ranges from 5\% of the sound speed at
the midplane to one-quarter the sound speed near the domain top and
bottom, as shown in figure~\ref{fig:coefft}.  Slower speeds within
0.5~AU of the boundaries are due to the restriction of the vertical
movement by the no-inflow condition.  Solutions of
equation~\ref{eqn:dampedwave} using the measured coefficients are
compared against the MHD spreading results in
figure~\ref{fig:allterms}.  Three cases are shown.  The common
practice in past chemical models is followed at top right, where the
second time derivative term is neglected and the diffusion equation is
solved with a uniform coefficient corresponding to the domain-averaged
total accretion stress.  At lower left, the diffusion coefficient
$\tau{\overline{v_z^2}}$ is allowed to vary with height.  At lower
right, all the terms are included in the damped wave equation.  The
damped wave solution best matches the MHD results.

\subsection{Accretion and Mixing}

\begin{figure}[tb]
\epsscale{0.4}
\plotone{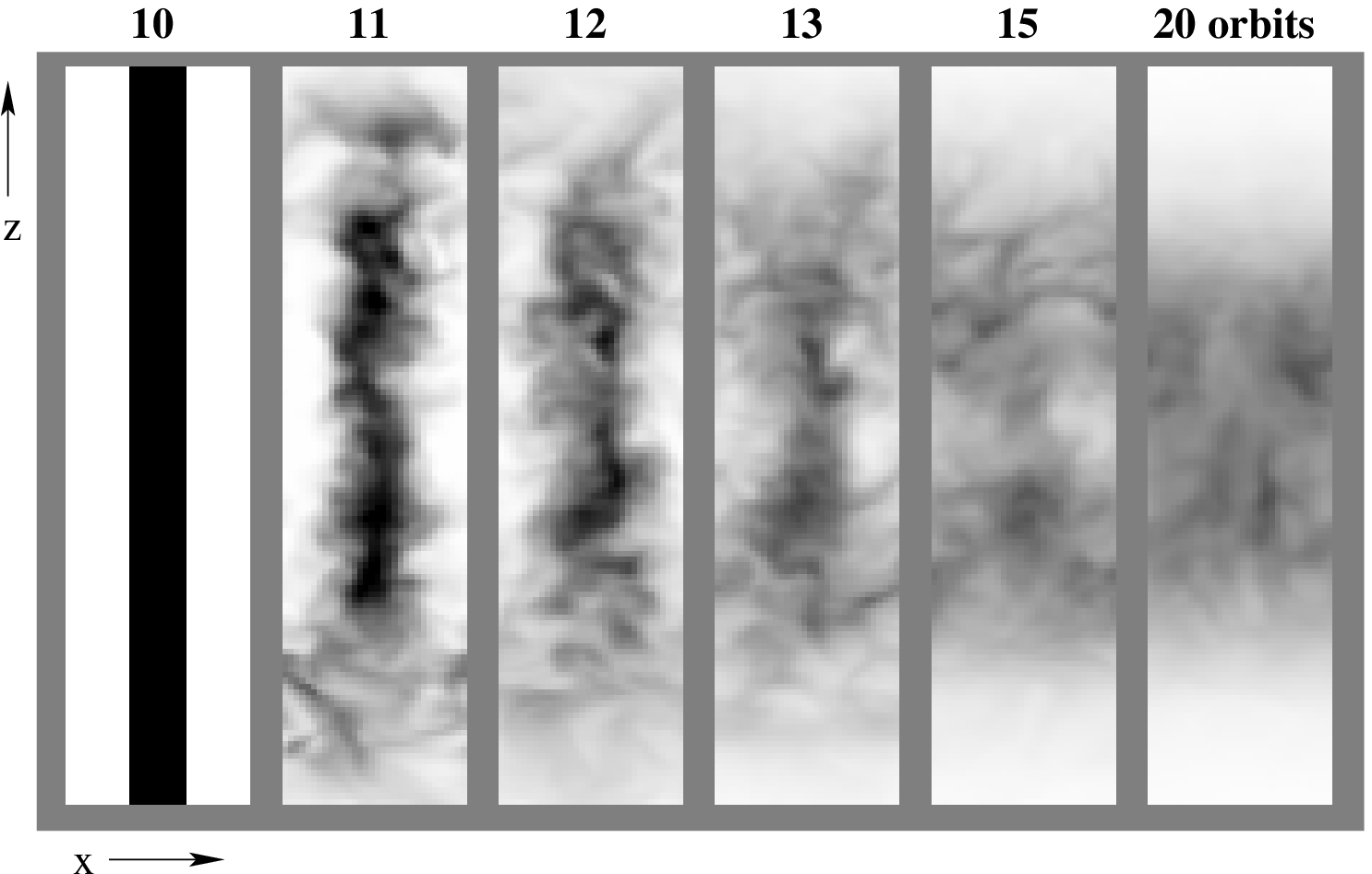}
\caption{\small Spreading of a contaminant in the radial and vertical
directions.  A uniform layer perpendicular to the $x$-direction is
placed in the turbulence at ten orbits.  Snapshots of the
azimuthally-averaged contaminant density are shown at six times.  The
fastest spreading is in the radial direction in the surface layers.
After five orbits the remaining radial gradients are small.  The
vertical mixing and the compression of gas moving toward the midplane
become the main effects.  The gray scale is linear between zero
(white) and the initial value (black).
\label{fig:radialmixing}}
\end{figure}

To find whether the mixing driven by the MRI is faster or slower than
the accretion, we compare the turbulent diffusion coefficients for
mixing in the radial and vertical directions with the angular momentum
transfer coefficient or kinematic viscosity corresponding to the
stress,
\begin{equation}
\nu = {w_{xy} \over \rho\left|{\partial\Omega/\partial\ln R}\right|}.
\end{equation}
The radial mixing is examined using a version of the fiducial run with
the contaminant placed initially in the middle third of the domain
width, and distributed uniformly in height and along the orbit.
Radial and vertical mixing occur together because the disk is
stratified.  As shown in figure~\ref{fig:radialmixing}, the mixing is
quickest in the surface layers.

\begin{figure}[tb]
\epsscale{0.4}
\plotone{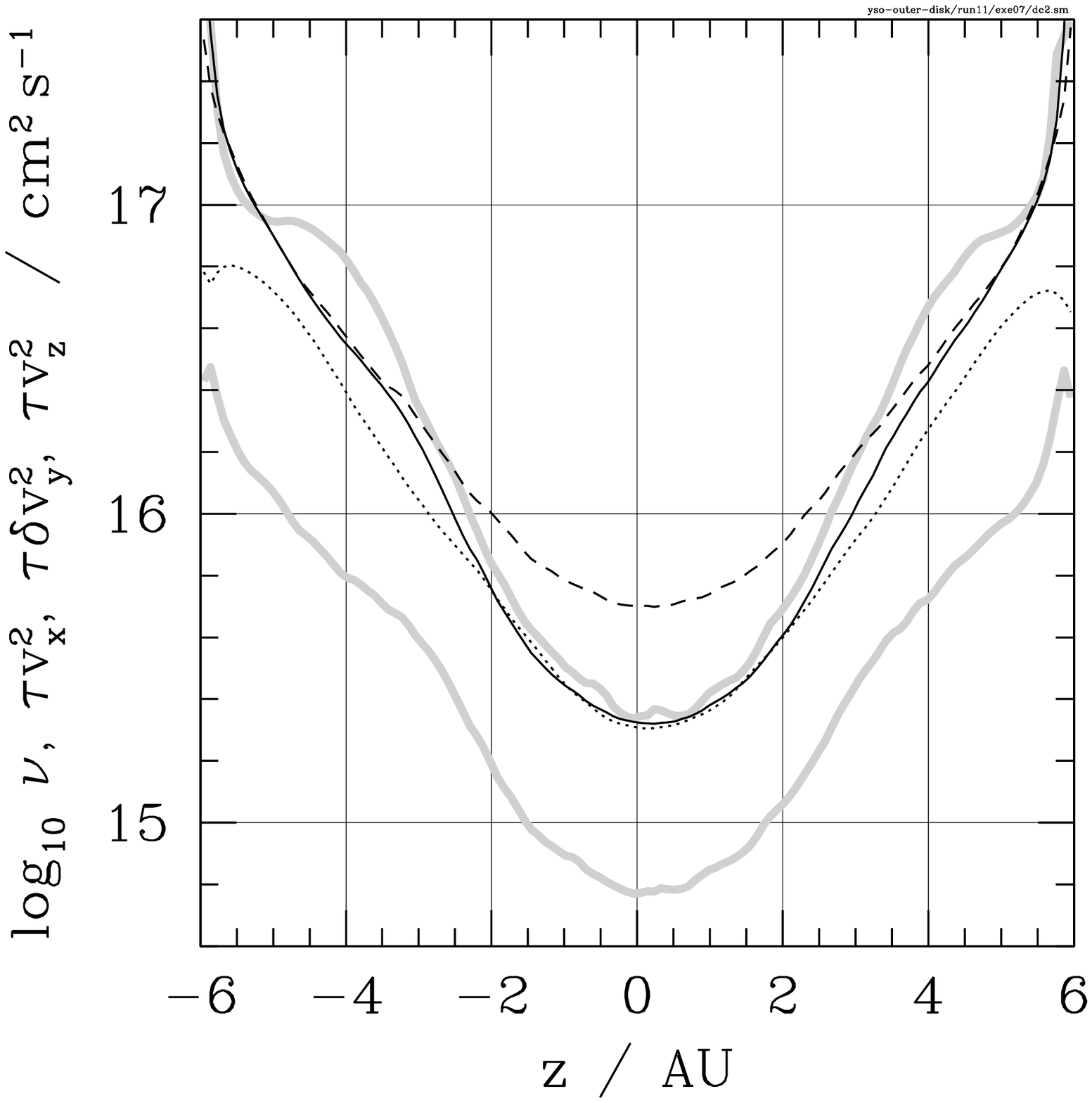}
\caption{\small Time- and horizontally-averaged angular momentum
transfer and turbulent mixing coefficients versus height.  Thick gray
lines show the total kinematic viscosity corresponding to the
accretion stress, $(\frac{3}{2}\Omega)^{-1} \langle
-B_xB_y/4\pi\rho+v_x \delta v_y\rangle$ (upper), and the hydrodynamic
part alone (lower).  Thin black curves mark the turbulent diffusion
coefficients for mixing in the radial (dashed), azimuthal (solid) and
vertical directions (dotted).  A coefficient $10^{16}$~cm$^2$s$^{-1}$
corresponds to a radial mixing time of 0.64~Myr.
\label{fig:dc}}
\end{figure}

The turbulent mixing and angular momentum transfer coefficients are
plotted against height in figure~\ref{fig:dc}.  The results from the
fiducial run have been averaged from 10 to 90~orbits, and the
turbulent correlation times for the radial and azimuthal mixing are
set equal to that measured for the vertical mixing.  There is an
overall height variation, with both accretion and mixing fastest in
the outer layers of the disk where the velocity dispersion is
greatest.  The accretion time $R^2/\nu$ of three million years at the
midplane is similar to the duration of the disk accretion phase in
star formation \citep{ss89,se93,hl01,mh04}, while the accretion time
at a height of 4~AU is just $10^5$~yr.

Accretion and vertical mixing occur at similar rates near the
midplane, while at heights of 4~AU the coefficient for radial angular
momentum transfer is about twice that for vertical diffusion.  The
radial mixing is faster than the vertical at all heights, in agreement
with the results shown in figure~\ref{fig:radialmixing} and also
consistent with the relative sizes of the velocity fluctuation
components in the calculations by \cite{ms00}.  The importance of
uncorrelated velocity fluctuations for the mixing is shown by the
large ratios of the diffusion coefficients to the kinematic viscosity
corresponding to the hydrodynamic part of the accretion stress.  The
hydrodynamic stress results only from correlated fluctuations in the
radial and azimuthal velocities.

\subsection{Effects of the Spatial Resolution and Boundaries
\label{sec:boundaries}}

The dependence of the results on the grid resolution is explored using
a version of the fiducial run with the number of zones along each axis
doubled to $64\times 128\times 256$.  Over twenty orbits, the time-
and horizontally-averaged profiles of the accretion stress and mixing
coefficient differ from the fiducial version by less than the range of
time variation.  Any sensitivity to the grid spacing is weak.

Possible effects of the vertical boundaries are checked by repeating
the first thirty orbits of the fiducial calculation in a taller
domain.  The height and the number of grid zones in the vertical
direction are increased by 25\%, to 7.5~AU either side of the midplane
and 160~zones.  The resulting vertical profiles of the density,
accretion stress, and mixing coefficient are similar where the two
calculations overlap, except within 0.7~AU of the top and bottom
boundaries.  The time-averaged accretion stress at the boundaries of
the fiducial calculation is up to twice that at the same height in the
taller version, due to magnetic forces where field lines exit the
domain.  The mixing coefficient in the taller calculation increases
smoothly with height through 6~AU, while in the fiducial run the
mixing coefficient declines within 0.7~AU of the boundaries.  We
conclude that the mixing rates are affected by the boundaries only in
the top and bottom 6\% of the domain height.  Solutions of
equation~\ref{eqn:dampedwave} without the boundary effects can be
obtained by projecting the approximately cubic height dependence of
the diffusion coefficient from nearer the midplane.

The density floor is generally applied only in a few zones near the
vertical boundaries.  The application of the floor in the fiducial run
between 10~and 30~orbits increases the mean density by less than one
part in $10^7$.  As a further check, a version of the taller
calculation with the floor reduced tenfold to $10^{-19}$ g~cm$^{-3}$
is run to twenty orbits.  The mean profiles of the pressure, accretion
stress and mixing coefficient are not significantly different from
those above.

The effects of the radial box size are tested by repeating the first
sixty orbits of the fiducial calculation in a box with the width and
the number of grid zones in the radial direction doubled, so that
there are $64\times 64\times 128$~zones.  The horizontally-averaged
accretion stress and velocity fluctuations are similar to those in the
fiducial run.

\subsection{Vertical Magnetic Fields}

The calculations described above have zero net vertical magnetic flux.
A mean vertical field leads to larger accretion stresses \citep{hg95}.
The flow is dominated by the repeated formation and break-up of
approximately axisymmetric pairs of inward and outward-moving layers,
in unstratified shearing-box calculations with the MRI wavelength a
few times shorter than the domain height \citep{si01}, while in
stratified calculations, the strong magnetic fields in the layers
disrupt the disk \citep{ms00}.  A similar disruption occurs in a
version of our fiducial run with the same uniform initial magnetic
pressure but the field lines vertical throughout.  The MRI reaches
non-linear amplitudes at 2.5~orbits, strong radial and azimuthal
fields are generated, the mean magnetic pressure exceeds the gas
pressure at 2.9~orbits, the ram pressure of the velocity fluctuations
is greater than the gas pressure after 3.6~orbits, and the calculation
is halted as mass and magnetic fields are ejected rapidly through the
top and bottom boundaries.  The shearing-box approximation is unsuited
to study vertical mixing in this situation.  Disk annuli threaded by
net vertical magnetic fields will be violently active and may be
short-lived, if the MRI wavelength is similar to the disk thickness.

\section{STEADY-STATE AND FLUCTUATIONS\label{sec:steady}}

In the final set of calculations we examine the distribution of the
contaminant in the presence of fixed sources.  The release of a
molecular species into the gas phase in the disk surface layers is
represented by holding the contaminant density fixed at its initial
value in the top and bottom ten percent of the domain height.  Two
calculations are carried out starting at ten orbits, after the
turbulence is well-established.  In the first, the concentration is
initially zero except in the surface-layer sources.  The contaminant
spreads gradually into the interior through the turbulent motions, and
the midplane concentration increases with time.  The compression of
the gas moving from the surface layers to the midplane leads to
midplane contaminant densities greater than the density at the source.
The gas density at the midplane averages 59 times greater than at the
inner edge of the source region, while after eighty orbits of mixing,
the contaminant density at the midplane is 5.4~times that at the
source and still increasing.  The time for the interior to reach
equilibrium with the source is clearly greater than the run length of
$13\,000$~years.  A rough estimate of the equilibration time is
obtained using the distance from source to midplane and the turbulent
diffusion coefficient at the midplane, where the mixing is slowest
(figure~\ref{fig:dc}).  The result, $(4.8 {\rm AU})^2 / (\tau
\overline{u_z^2}[z=0]) = 80\,000$~yr, is about six times the run
duration.

\begin{figure}[tb]
\epsscale{0.4}
\plotone{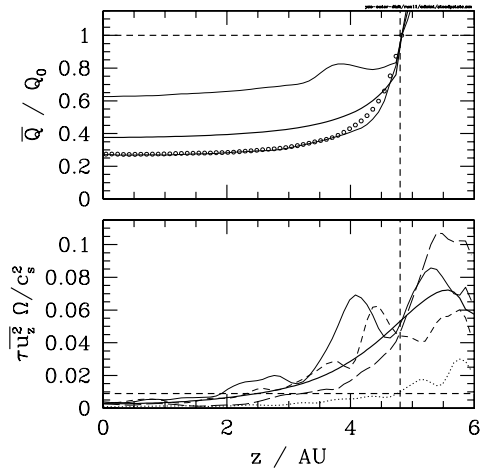}
\caption{\small The vertical profile of contaminant concentration (top
panel), averaged over horizontal planes and over time (thick solid
line), in the calculation starting with the steady-state solution of
the damped wave equation~\ref{eqn:dampedwave}.  The initial state is
marked by circles and the time variation is shown by the profiles at
the times of lowest and highest midplane density, 10.4 and
54.6~orbits, respectively (thin solid lines).  A vertical dashed line
lies at the boundary of the source region where the contaminant
density is fixed, and a horizontal dashed line shows the concentration
at the source.  In the bottom panel the diffusion coefficient is
plotted against height (thick solid line).  The coefficient is taken
from the previous MHD calculation and is averaged horizontally as well
as between the top and bottom halves of the domain and over time.
Also shown are the horizontally-averaged profiles at 20, 40, 60 and
80~orbits (thin solid, dotted, dashed and long-dashed curves,
respectively) together with the kinematic viscosity (thick horizontal
dashed line) computed from the domain-averaged total accretion stress
and the domain-averaged gas density.
\label{fig:steadystate}}
\end{figure}

In the second calculation with fixed sources, the contaminant density
in the disk interior is initially set to the steady-state solution of
the damped wave equation~\ref{eqn:dampedwave}.  The damped wave and
diffusion equations have the same equilibrium solution since they
differ only by a time-derivative term.  The diffusion coefficient is
obtained by averaging horizontally and over time in the previous MHD
calculation.  The horizontally-averaged profile of the contaminant
remains near the steady-state solution for eighty orbits despite
fluctuations in the mixing rate, as shown in
figure~\ref{fig:steadystate}.  The midplane contaminant density varies
around its starting point, with the largest value just 8\% greater
than the smallest, and the time average 25~times that at the source.
The equilibrium contaminant concentration is least at the midplane,
where the mixing is slowest.

\begin{figure}[tb]
\epsscale{0.4}
\plotone{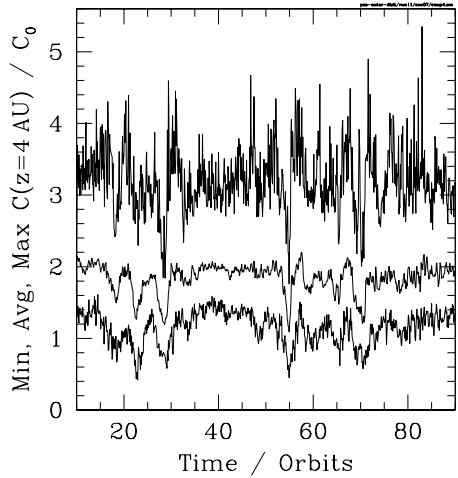}
\caption{\small Maximum, average and minimum contaminant densities on
the horizontal surface 4~AU above the midplane, as functions of time
in the MHD calculation near steady-state.  The contrast is typically
about a factor three.  The densities are measured relative to the
value in the source layers.
\label{fig:xyfluctuations}}
\end{figure}

While the horizontally-averaged profile remains near a steady state,
large horizontal fluctuations in the contaminant density persist
throughout the run, due to changes in the rate and direction of the
turbulent transport.  The largest contaminant density at a height of
4~AU is two to seven times greater than the smallest.  The density
range and its time variations are shown in
figure~\ref{fig:xyfluctuations}.  The variations occur over timescales
from the turbulent correlation time up to the duration of the
calculation.  Variations of this kind are missing from hydrostatic
models and are large enough to be important for the overall chemical
evolution of protostellar disks.

\begin{figure}[tb]
\epsscale{0.4}
\plotone{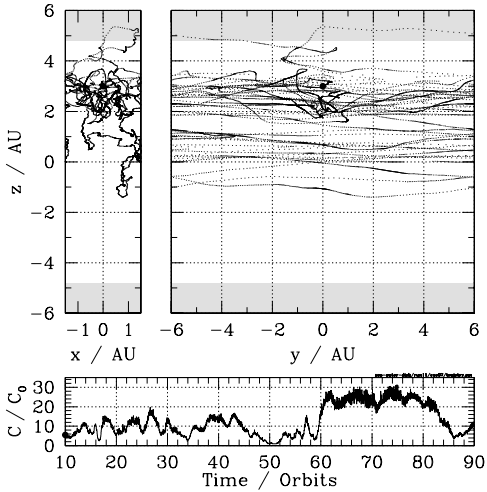}
\caption{\small Trajectory of a fluid element in the turbulence,
projected onto the $x-z$ plane (top left) and the $y-z$ plane (top
right).  The element lies 3~AU above the origin at 10~orbits (filled
circles) and follows a path that takes it into the upper contaminant
source layer (gray) and across the midplane, and at 90~orbits (open
circles) returns near the original location.  The time history of the
local contaminant density is plotted in the bottom panel.
\label{fig:trajctry}}
\end{figure}

Individual fluid elements in the disk may pass through a wide range of
environments on their paths through the turbulence.  The chemical
compositions of the fluid elements will reflect the histories of local
conditions if the chemical reactions fail to reach equilibrium over
the flow timescales.  The history of a representative fluid element in
the MHD calculation with the contaminant near steady-state is shown in
figure~\ref{fig:trajctry}.  Over $13\,000$~years the element traverses
the whole range of heights.  The ambient contaminant density varies by
around 20\% on the turbulent correlation timescale of 30~years, and by
factors of three over a few hundred years due to horizontal motions
through fluctuations like those shown in
figure~\ref{fig:xyfluctuations}.  The density occasionally changes by
an order of magnitude within a thousand years as the fluid element
moves rapidly in height.  Note that the path as viewed along the
$x$-direction in the top right panel of figure~\ref{fig:trajctry} is
broken where the element crosses the shearing-periodic radial
boundary, leading to an offset along the $y$-direction.  Paths like
that in figure~\ref{fig:trajctry} will also be traced by dust grains
that are well-coupled to the gas by drag forces.

\section{SUMMARY AND CONCLUSIONS\label{sec:conc}}

We have followed the mixing of a passive trace species and the
trajectories of individual fluid elements in the outer Solar nebula,
using 3-D MHD calculations.  A small patch of the nebula centered at a
radius of 30~AU is modeled in the shearing-box approximation, with the
vertical component of the gravity of the young Sun included.  The
mixing is caused by the turbulence driven by the magneto-rotational
instability.  The magnetic and hydrodynamic stresses in the turbulence
correspond to a kinematic accretion viscosity that is greatest in the
surface layers because of (1) the density stratification and (2) the
stronger magnetic fields in the surface layers.  The mixing along the
vertical direction is well-matched by the solutions of a
one-dimensional damped wave equation, with damping coefficient
approximately the product of the shear time and the mean squared
vertical turbulent speed.  A similar horizontally-averaged diffusion
description of the mixing without the wave term gives a poorer fit to
the MHD results: the contaminant spreads faster than the RMS turbulent
speed at early times, owing to the infinite signal propagation speeds
that occur generically in solutions of the diffusion equation.  The
horizontal averaging used in both the damped wave and diffusion
pictures glosses over some details of the flow that could affect the
chemical evolution.  The contaminant density in the MHD calculations
varies substantially across horizontal planes at any given time, due
to fluctuations in the speed and direction of the gas flows.  The
mixing caused by the turbulence is erratic, and can be expected to
intermittently force the nebula away from local chemical equilibrium.
Also, individual fluid elements sometimes move between the interior
and the surface layers in a small fraction of the time needed to reach
an overall mixing equilibrium.  However an improvement over existing
one-dimensional treatments of vertical mixing can be obtained by
solving equation~\ref{eqn:dampedwave}.

The measured vertical mixing coefficient varies with height in a
similar way to the angular momentum transfer coefficient.  The two are
equal at the midplane, while at 4~AU the mixing coefficient is about
half the local kinematic viscosity, so that the Schmidt number ranges
from one to two.  The accretion time in the midplane is three million
years, comparable to the lifetimes of the disks around T~Tauri stars.
The accretion time in the surface layers is less than $10^5$~yr, and
similar to the timescale for vertical mixing to the midplane.  The
transport times suggest a picture in which the fastest way to move
midplane material from one radius to another is by first mixing it out
to the surface layers.  In this scenario, only the solid particles
small enough to be suspended in the low-density surface-layer gas are
carried efficiently.  Larger bodies remain near the midplane where
both accretion and mixing are slow.  Such fast surface-layer transfer
could lead to size-dependent composition differences among primordial
dust grains and could affect the molecular abundances in the region
where the giant planets formed.  The height dependence of the mixing
rate means that concentration gradients persist even in the steady
state.  The stratification of the turbulence may also be significant
for grain growth.  The RMS velocity fluctuations increase with height,
exceeding 100~m~s$^{-1}$ above 4~AU.  Aggregate icy grains can be
broken apart in impacts at these speeds \citep{dt97}, while bodies
that settle near the midplane will experience mostly slower collisions
leading to compaction or sticking.  The vertical mixing may lie behind
the difference in HCN deuterium fraction between comets and the
interstellar medium.  Molecules formed in the surface layers of the
nebula, where warm temperatures led to lower deuteration, could have
been incorporated into bodies assembled near the midplane.  There is a
need for grain growth and chemical evolution calculations that include
the effects of both radial and vertical transport.

\begin{acknowledgments}
This work was supported by the National Research Council through a
fellowship to NJT, and by the National Aeronautics and Space
Administration through an Origins of Solar Systems program grant to KW
and a Terrestrial Planet Finder Foundation Science program grant to GB
and HWY.  The supercomputer used for the high-resolution calculation
was funded by the Jet Propulsion Laboratory, Institutional Computing
and Information Services, and the NASA Directorates of Aeronautics
Research, Science, Exploration Systems, and Space Operations.  JPL is
operated by the California Institute of Technology under contract to
NASA.
\end{acknowledgments}


\end{document}